\documentclass[a4paper,12pt]{article}

\usepackage{geometry,slashed}
 \usepackage
 {
   amsmath,amssymb,graphicx,tabularx
 }
  \usepackage{soul}
  \usepackage[usenames,dvipsnames]{xcolor}
 
 \definecolor{X575}{rgb}{0.05, 0.7, 0.05}
 
 \usepackage{color}
 \usepackage{hyperref}
 \usepackage{caption,subcaption}
 \usepackage{authblk}
 \usepackage{epstopdf}
 \usepackage{enumerate}
 \usepackage{multirow}
 \usepackage{siunitx}
 \usepackage{float}
 \usepackage{booktabs}
 \usepackage{cite}

  \usepackage{listings}
  \usepackage{calrsfs}
  \DeclareMathAlphabet{\pazocal}{OMS}{zplm}{m}{n}

\numberwithin{equation}{section}

\title{Subleading EW corrections and spin-correlation effects in $t\bar{t}W$ multi-lepton signatures}

\author[1]{Rikkert Frederix\thanks{rikkert.frederix@thep.lu.se}}
\author[2]{Ioannis Tsinikos\thanks{ioannis.tsinikos@thep.lu.se}}

\affil[1,2]{\small Theoretical Particle Physics, Department of Astronomy and Theoretical Physics, Lund University, S\"olvegatan 14A, SE-223 62 Lund, Sweden}

\begin{document}

\maketitle

\vspace*{-9cm}
\noindent
{\small LU-TP~20-19}
\vspace*{9cm}

\begin{abstract}
  Recently a slight tension between data and predictions has been
  reported in $t\bar{t}W$ production by both the CMS and ATLAS
  collaborations. We revisit the theoretical predictions for
  this process, focussing on the following two effects. We
  disentangle various effects that lead to asymmetries among the
  leptonic decay products of the \mbox{(anti-)top} quarks and $W$
  bosons, for which we find that the spin correlations in the top-quark 
  pair are the dominant source. We also discuss the impact
  of the large, formally subleading, electroweak corrections to
  $t\bar{t}W$ production at the LHC. We find that this effect changes
  the $t\bar{t}W$ cross section significantly in the signature
  phase-space regions, and should therefore be included differentially
  in the theory to data comparisons.
\end{abstract}

\section{Introduction}
\label{sec:intro}

With the 13 TeV LHC run, both ATLAS and CMS collaborations have
measured the $t \bar{t} V (V=Z,W)$ cross sections. These processes are
studied either independently~\cite{Aaboud:2019njj,Sirunyan:2017uzs} or
as irreducible backgrounds to $t \bar{t} H$ (multilepton)
searches~\cite{ATLAS:2019nvo,CMS:2017vru}. In both cases a slight
tension between theoretical predictions and data is observed for $t
\bar{t}W$ production, with the data suggesting a somewhat larger cross
section than Standard Model predictions. This slight tension between
Standard Model predictions and data warrants further study of this
process from both the experimental and the theoretical sides.

At the production level the $t \bar{t}W$ process has recently been studied
in detail at the complete-NLO accuracy~\cite{Frederix:2017wme}. In
this work it was pointed out that the subleading EW corrections
result in a $\sim$10\% increase of the total cross section. This large
contribution is due to the opening of $tW \to tW$ scattering
diagrams. The complete-NLO calculation has been matched to soft
(threshold) gluon resummation, resulting in the most-accurate
predictions for the $t \bar{t}W$ production at the LHC to
date~\cite{Broggio:2019ewu,Kulesza:2020nfh}. Both these works have
shown that $t \bar{t}W$, in contrast to $t \bar{t}Z$ and $t \bar{t}H$,
does not become less sensitive to the scale choices even when
including the resummation at NNLL accuracy. In other words, including
the all-order soft-gluon resummation does not significantly decrease
the theoretical uncertainties. This can predominantly be attributed to
the absence of gluon-induced channels at LO; the latter only appear at
higher orders and give sizeable contributions to the cross
section. Since they do not contribute at LO they are not considered in
the resummation frameworks of
refs.~\cite{Broggio:2019ewu,Kulesza:2020nfh}. Incidentally, since the
$gg$ channels only open at NNLO accuracy there is a large $t \bar t$
asymmetry of $\sim$3\% in $t\bar{t}W$
production~\cite{Maltoni:2014zpa,Maltoni:2015ena,Broggio:2019ewu}.

At the decay level this accuracy cannot be maintained due to the
complexity of the calculation. It is shown in
ref.~\cite{Maltoni:2014zpa} that the presence of the $W$ boson 
polarises the initial quark-line and in turn the final $t \bar{t}$
pair. The emerged lepton asymmetry is of $\sim$-13\% at NLO in QCD and
has consequences on the final lepton pseudo-rapidity distributions. 
Therefore it affects the fiducial region of the final multi-lepton
signatures depending on the applied cuts. Furthermore the subleading
EW $tW \to tW$ scattering contributions are governed by different
kinematics and as a result a non-flat effect is expected in the
fiducial region.

Since the largest tension for $t \bar{t}W$ is found when it enters as
the main background in the $t \bar{t}H$ multi-lepton analyses, these
multi-lepton signatures is what we focus on in this work. We study all
effects that lead to lepton asymmetries in detail. We further study
contributions that have not been yet taken under consideration in the
fiducial region: for the first time, we include the subleading EW
corrections in a consistent matching to the parton shower. The latter
allows us to investigate the effects from these large corrections
differentially in the fiducial region relevant to the $t\bar{t}H$
multi-lepton signatures.

The structure of this paper is the following: in
section~\ref{sec:calc_par} we discuss the input parameters, describe
the framework of the calculation and we define the experimental
fiducial region under study. In section~\ref{sec:results} we discuss
the main subleading EW and spin-correlation effects on differential
distributions and their impact on measurements. We present our
conclusions in section~\ref{sec:concl}.

\section{Input parameters and calculation setup}
\label{sec:calc_par}
We consider the NLO corrections to both $pp \to t\bar{t}W^+$ and $pp
\to t\bar{t}W^-$ production in the fiducial region following the ATLAS
analysis of ref.~\cite{ATLAS:2019nvo}. The calculation is performed
within the MadGraph5\_aMC@NLO~\cite{Alwall:2014hca} framework including
the automation of the EW calculations~\cite{Frederix:2018nkq}. In
accordance with the notation of \cite{Frederix:2017wme} we define for
any observable the QCD and subleading EW (EW$_{\rm sub}$)
perturbative orders as following:
\begin{align}
\Sigma_{\rm QCD} &= \alpha_s^2 \alpha \Sigma^{t \bar t W}_{3,0} + \alpha_s^3 \alpha \Sigma^{t \bar t W}_{4,0} \nonumber \\
                &= \Sigma_{\textrm{LO}_1} + \Sigma_{\textrm{NLO}_1} \nonumber \\
\Sigma_{{\rm EW}_{\rm sub}} &= \alpha^3 \Sigma^{t \bar t W}_{3,2} + \alpha_s \alpha^3 \Sigma^{t \bar t W}_{4,2} \nonumber \\
                &= \Sigma_{\textrm{LO}_3} + \Sigma_{\textrm{NLO}_3} \,. 
\label{eq:blobs}
\end{align}
We perform the calculation in the 5 Flavour Scheme, setting the
factorisation and renormalisation scales to $\mu=\frac{H_T}{2}$ and
using the NLO PDF4LHC PDF sets, with associated value for the strong
coupling. As input parameters we use
\begin{align}
m_t &= 173.34 \; {\rm GeV} \; , \; m_Z = 91.1876\; {\rm GeV} \nonumber \\
a_{EW} &= 132.232 \; , \; G_f = 1.16639 \times 10^{-5}\,.
\end{align}
The top quarks are decayed to $b$ quarks and $W$-bosons with a
branching ratio of 100\%. The $W$ bosons are decayed inclusively,
i.e.~both the prompt $W$ bosons and the ones induced by the top quark
decays are allowed to decay to quarks and leptons. Unless stated
otherwise, all these decays are realised within the MadSpin
framework~\cite{Artoisenet:2012st} in order to fully keep the (LO)
spin correlations.

We match the calculation to the parton shower using the PYTHIA8
framework~\cite{Sjostrand:2014zea} in the default tune, using
MadGraph5\_aMC@NLO's build-in MC@NLO matching technique
\cite{Frixione:2002ik,Frixione:2003ei}.
The reason that the EW$_{\rm sub}$ contribution can be included in the
matching to the parton shower is that the perturbative order $\alpha_s
\alpha^2$ (the LO${}_2$ in the notation of
ref.~\cite{Frederix:2017wme}) is exactly zero for this process. As a
result the $\alpha_s \alpha^3$ order can be considered as pure QCD
corrections to the $\alpha^3$ one\footnote{What are usually called the
  EW-corrections, i.e.~the $\alpha_s^2 \alpha^2$ perturbative order
  (a.k.a.~NLO${}_2$), are not included here. These EW corrections change
  the cross section by about
  $\sim$-4\%~\cite{Frixione:2015zaa,Frederix:2018nkq}.}.

In order to understand the spin-related and the subleading EW
effects, before applying any particle selection or cuts, we define the
inclusive (no cuts) signature. Furthermore, once specified, we select
only events for which the top-quark pair decays to a muon pair and the
associated $W$ boson to an electron(positron), using MC-truth.  This is
done only in order to pin down the origin of various effects and for
our final results the decays are inclusive. For the signal-region
definitions we start with the selection and the cuts, for which we
follow the experimental analysis of \cite{ATLAS:2019nvo}. We identify
the particles as following:
\begin{align}
{\rm Electrons:}\,\, &\; p_T(e) \ge 10 \; {\rm GeV} \; , \; |\eta(e)| \le 2.47 \; (2 \; {\rm for} \; {\rm tight}) \nonumber \\
{\rm Muons:}\,\, &\; p_T(\mu) \ge 10 \; {\rm GeV} \; , \; |\eta(\mu)| \le 2.5 \; ({\rm same} \; {\rm for} \; {\rm tight}) \nonumber \\
{\rm jets:}\,\, &\; k_T =-1 \; , \; R=0.4 \; , \; pT(j) \ge 25 \; {\rm GeV} \; , \; |\eta(j)| \le 2.5 \,.
\label{eq:sel}
\end{align}
The $\tau$ leptons are allowed to decay within the shower and we
identify the hadronic $\tau_h$. We reject the jets that have $\Delta R
(j,e) \le 0.3$ or $\Delta R (j,\tau_h) \le 0.3$. Furthermore we
discard the muons that lie within $\Delta R (j,\mu) \le 0.4$.

With this particle selection we define the two following signatures: the
same sign dilepton (2ss$\ell$) and the three lepton (3$\ell$)
channels. In both cases we require at least two jets and at least one
b-tagged jet and zero hadronic $\tau_h$'s. For the 2ss$\ell$ signature
we require exactly two tight same sign leptons with $p_T(\ell) \ge 20$
GeV. Furthermore for the same flavor (SF) pairs we apply the $m(\ell
\ell) \ge 12$ GeV condition. For the 3$\ell$ signature we ask exactly
three leptons, two tight same sign (SS) with $p_T(\ell) \ge 15$ GeV
and one opposite sign (OS) with $p_T(\ell) \ge 10$ GeV. For the SFOS
pairs we ask $m(\ell^+ \ell^-) \ge 12$ GeV, $|m(\ell^+ \ell^-)-m_Z
|\ge 10$ GeV and for the 3-lepton system $|m(\ell \ell \ell)-m_Z |\ge
10$ GeV.

In PYTHIA8 we include hadronisation, the QED shower (we include the
photons in the jets) and the multiparton interactions (underlying
event). However, we do not consider any misidentification or
identification inefficiencies for the jets or the leptons.

\section{Results}
\label{sec:results}
Having defined the selection criteria for the particles and the events
we proceed now by pointing out the importance of the spin correlations
and thereafter showing the effect of the EW$_{\rm sub}$
contributions. In particular, we focus on the jet-multiplicity cross
sections, since that is shown by the ATLAS collaboration in
ref.~\cite{ATLAS:2019nvo} (both with prefit and postfit
signal+background contributions). However, since the data has not been
unfolded, we cannot directly compare to it. On the other hand, we can
study this distribution at the theoretical level to see firstly
whether and how it is shaped by the spin correlations and secondly if
the EW$_{\rm sub}$ effects considered in this work might have a
significant influence on the ATLAS analysis.

\subsection{Asymmetries}
\label{sec:Spin}
\begin{figure}[h]
\centering
\includegraphics[width=0.425\textwidth]{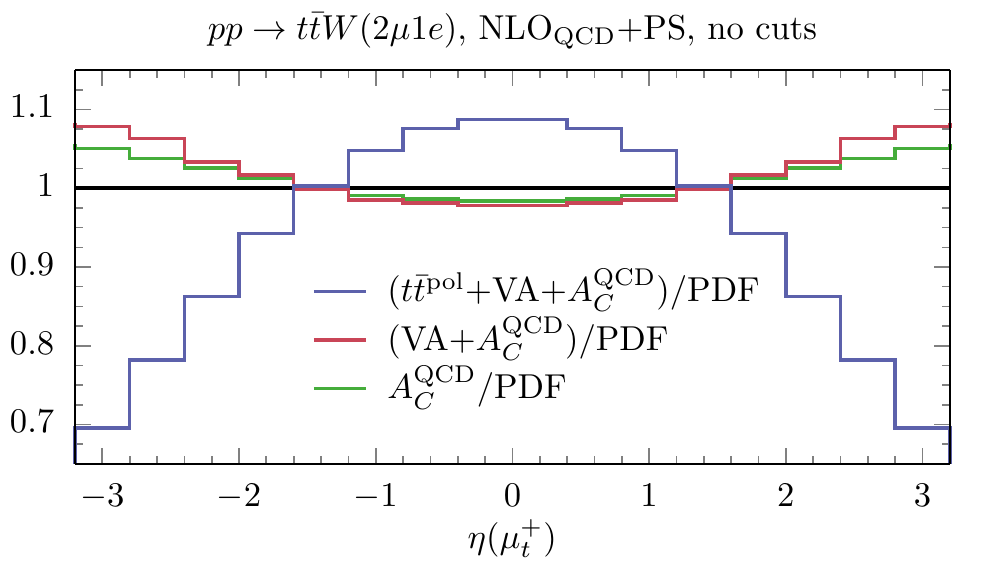}
\includegraphics[width=0.425\textwidth]{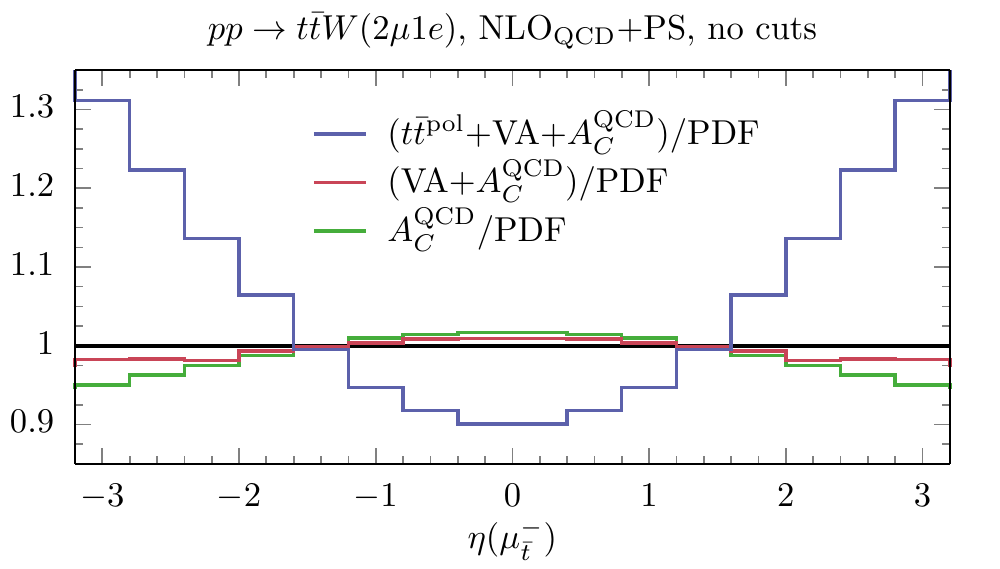} \\
\includegraphics[width=0.425\textwidth]{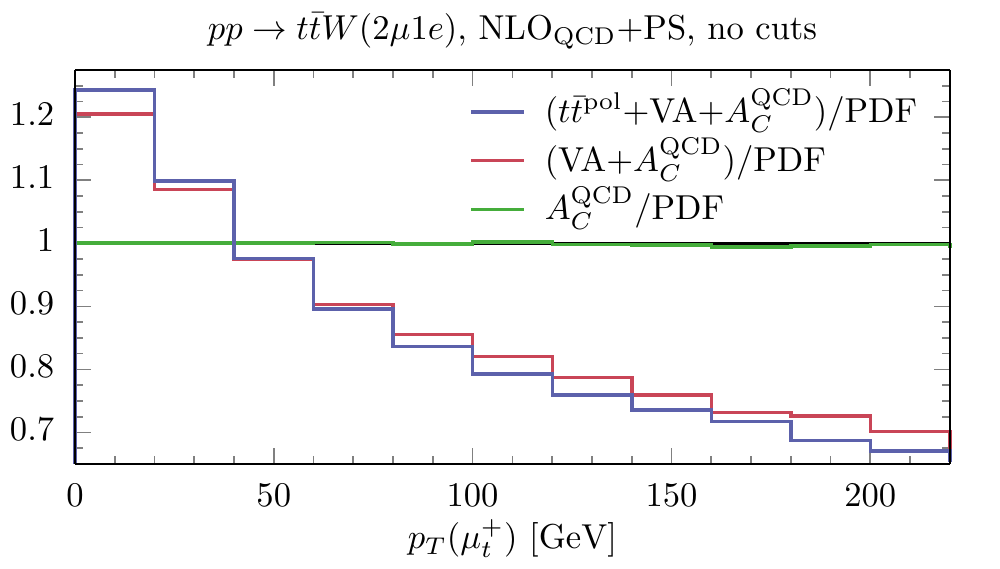}
\includegraphics[width=0.425\textwidth]{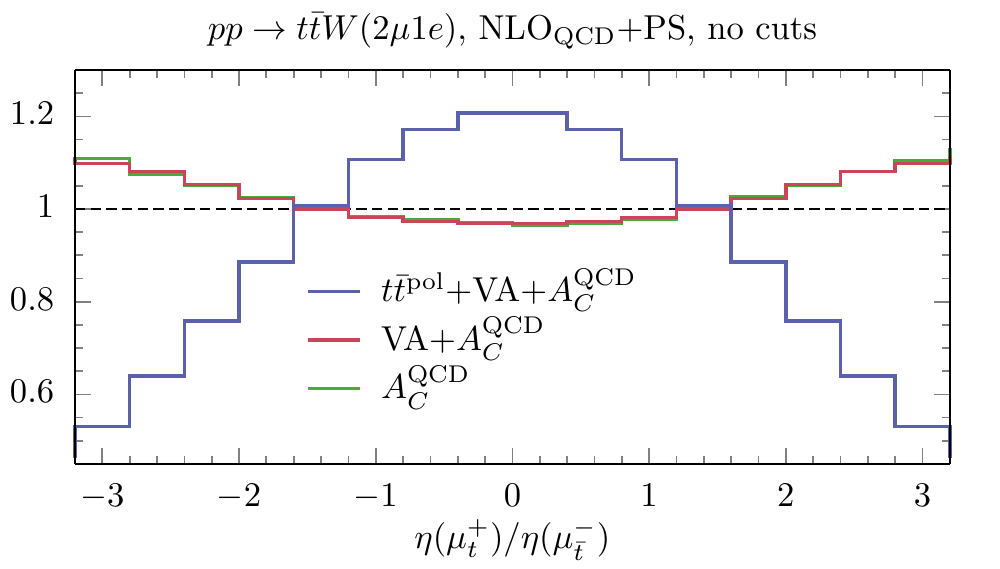}
\caption{Origin of asymmetries in top pair decay products. The muons in these plots exclusively originate from the top-quark pair.}
\label{fig:asymm}
\end{figure}

\begin{figure}[h]
\centering
\includegraphics[width=0.425\textwidth]{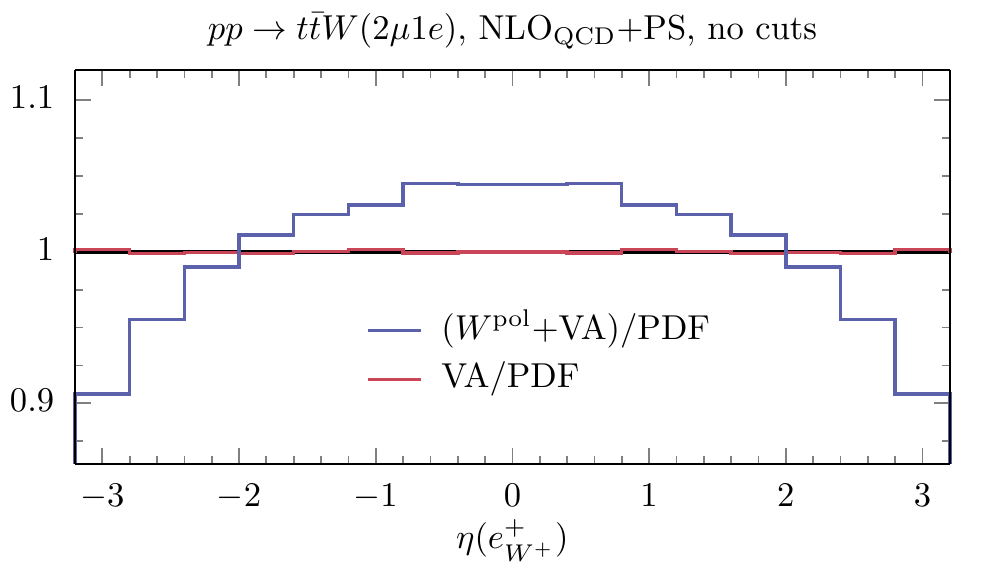}
\includegraphics[width=0.425\textwidth]{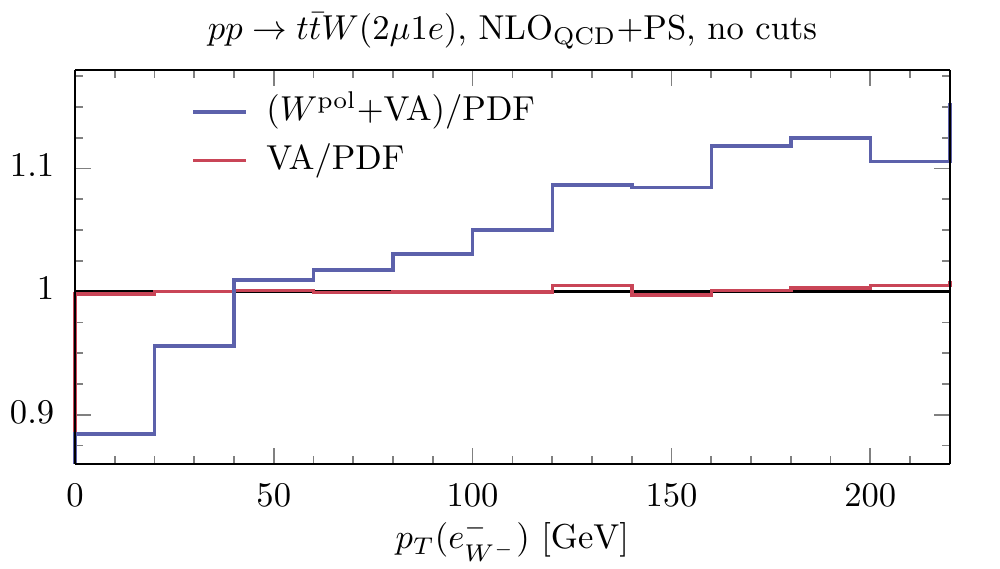} \\
\includegraphics[width=0.425\textwidth]{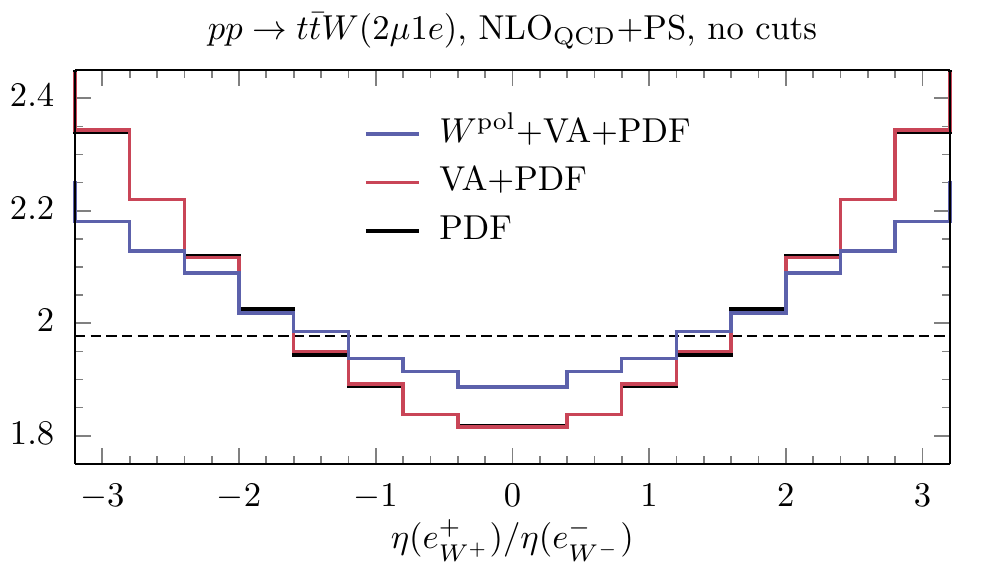}
\includegraphics[width=0.425\textwidth]{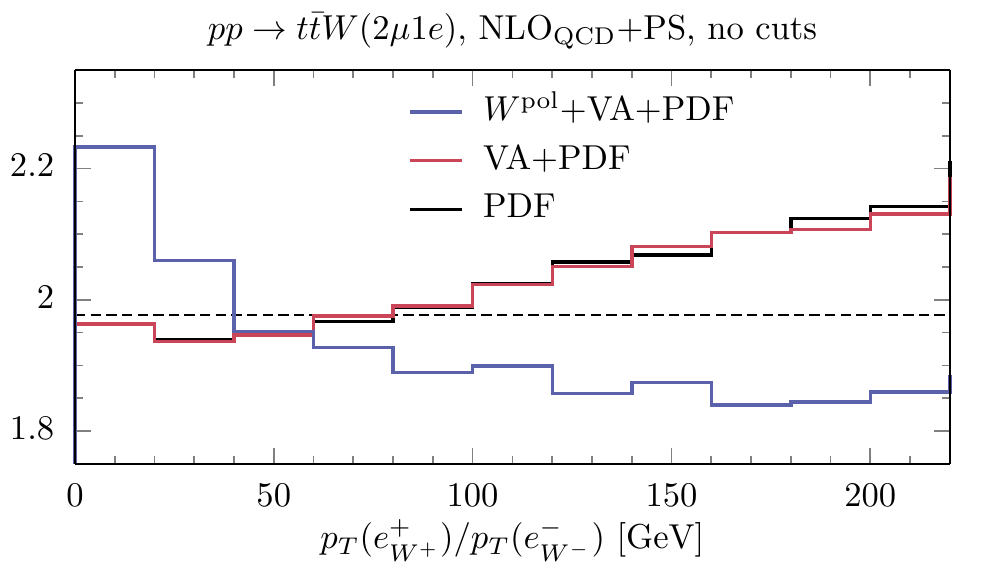}
\caption{Origin of asymmetries in $W$ associated decay products. The electrons(positrons) in these plots originate exclusively from the $W^-(W^+)$ associated boson.}
\label{fig:assoc}
\end{figure}

The effects described in the present section are already included in the modern MC simulations. 
Nevertheless, they are never disentangled in such detail in order to scrutinise their impact 
and understand their contribution to the final signatures.
Indeed, the asymmetries in the lepton decay products of the $t\bar{t}W^+$ and
$t\bar{t}W^-$ can be attributed to separate origins, which are depicted 
in figs.~\ref{fig:asymm} and \ref{fig:assoc} and described in what follows.
\noindent

For these plots, in order to track the origin of the leptons, we have
selected the events where the top quark pair decays to muons and the
associated $W^-(W^+)$ to electrons(positrons). Furthermore we restrict
the analysis only to the QCD shower within PYTHIA8 without applying
any cuts or selections. We denote the origin of each lepton with a
subscript. In fig.~\ref{fig:asymm} we show the various effects on the
decay products of the top pair, whereas in fig.~\ref{fig:assoc} of the
associated $W$ boson. We now separate these effects on the lepton
asymmetries:

\begin{itemize}
\item The $t\bar{t}W^+$ production is induced predominately by the
  $u\bar{d}+c\bar{s}$ luminosity, while $t\bar{t}W^-$ one mostly by
  $\bar{u}d+\bar{c}s$. This results in the total cross section for top
  pairs associated with the positively charged vector boson to be
  about a factor two larger than the negatively charged vector
  boson. Moreover, $t\bar{t}W^+$ typically probes larger Bj\"orken $x$
  values than $t\bar{t}W^-$ resulting, on average, in harder and more
  forward leptons in the former as compared to the latter.  This
  effect in figs.~\ref{fig:asymm} and \ref{fig:assoc} is defined as
  `PDF' and it affects both the top pair and the $W$ associated decay
  products. All the other effects are added on top of this. Since in
  fig.~\ref{fig:asymm} there is no distinction made for muons coming
  from $t\bar{t}W^+$ versus muons coming from $t\bar{t}W^-$, there is
  no PDF effect visible here. On the other hand, for the
  electron/positron coming from the associated $W$-boson decay this
  distinction is made, and therefore the PDF effect is clearly
  visible in the lower two plots of fig.~\ref{fig:assoc}, with the
  positron from the $t\bar{t}W^+$ process being at larger rapidities
  and harder than the electron from the $t\bar{t}W^-$ process.
\item The second cause for differences between the leptonic decay
  products is due to the Central-Peripheral asymmetry in the top
  pairs\cite{Kuhn:2011ri,Hollik:2011ps,Bernreuther:2012sx,Czakon:2014xsa,Czakon:2017lgo}. 
  This effect only enters at NLO in QCD, and was first studied
  for top pair production at the Tevatron, where it showed as a charge
  asymmetry~\cite{Aaltonen:2012it,CDF:2013gna,Abazov:2014cca,Aaltonen:2017efp}. 
  Compared to $pp\to t\bar{t}$ production, requiring the
  associated $W$-boson increases the asymmetry~\cite{Maltoni:2014zpa}. In fig.~\ref{fig:asymm}
  this is denoted as `$A_C^{\rm QCD}$'. It results in about 3-5\% differences 
  in the pseudo-rapidity of the lepton coming from the top decay versus the 
  one coming from the anti-top decay (pseudo-rapidity insets in fig.~\ref{fig:asymm}).
  It has a negligible effect on the corresponding $p_T$ distributions (fig.~\ref{fig:asymm}).
\item The third effect that could lead to asymmetries among the
  leptonic W-boson decays is the $V\!-\!A$ structure of the $W$-boson
  couplings. This is denoted in figs.~\ref{fig:asymm} and
  \ref{fig:assoc} as `VA' effects.  While these effects are not there
  for the decays of the associated $W$-boson (fig.~\ref{fig:assoc}),
  the $V\!-\!A$ coupling structure in the three-body decays of the top
  and the anti-top quarks results in a large asymmetry between the
  charged leptons and the neutrinos. They do affect the charged
  leptons from the top identically to the ones from the anti-top, and
  therefore do not generate an asymmetry in the visible lepton decays
  (fig.~\ref{fig:asymm}). Concerning the leptons originating from the
  associated $W$ boson, the `VA' effects are present once the spin
  correlations of the associated $W$ boson (denoted as $W^{\rm pol}$)
  are taken into account. In this case, as shown in
  fig.~\ref{fig:assoc}, the effects are different between the
  associated $W^+$ and $W^-$ and affect both the transverse momentum
  and pseudo-rapidity ratios.
\item The most important source for the asymmetry is the top-quark
  pair polarization. Due to the associated $W$, these correlations are
  rather sizeable and significantly alter the shapes of the rapidities
  of the leptonic decays of the top quark as compared to the leptonic
  decays of the anti-top quark (denoted as $t \bar t ^{\rm pol}$ in
  fig.~\ref{fig:asymm}). These effects result in a charge asymmetry of
  the leptons that reaches about -13--15\% at the cross section
  level~\cite{Maltoni:2014zpa}.  As shown in fig.~\ref{fig:asymm} in
  the pseudo-rapidity ratio these effects are between $\sim$+20\%
  (central region) and $\sim$-40\% (peripheral region).
\item The final source of asymmetry is due to the NLO electroweak
  corrections being of different size for $t\bar{t}W^+$ and
  $t\bar{t}W^-$ production. This is a negligible effect of
  $\sim$0.5\%~\cite{Broggio:2019ewu} and not considered 
  in this work.
\end{itemize}

\begin{figure}[t]
\centering
\includegraphics[width=0.425\textwidth]{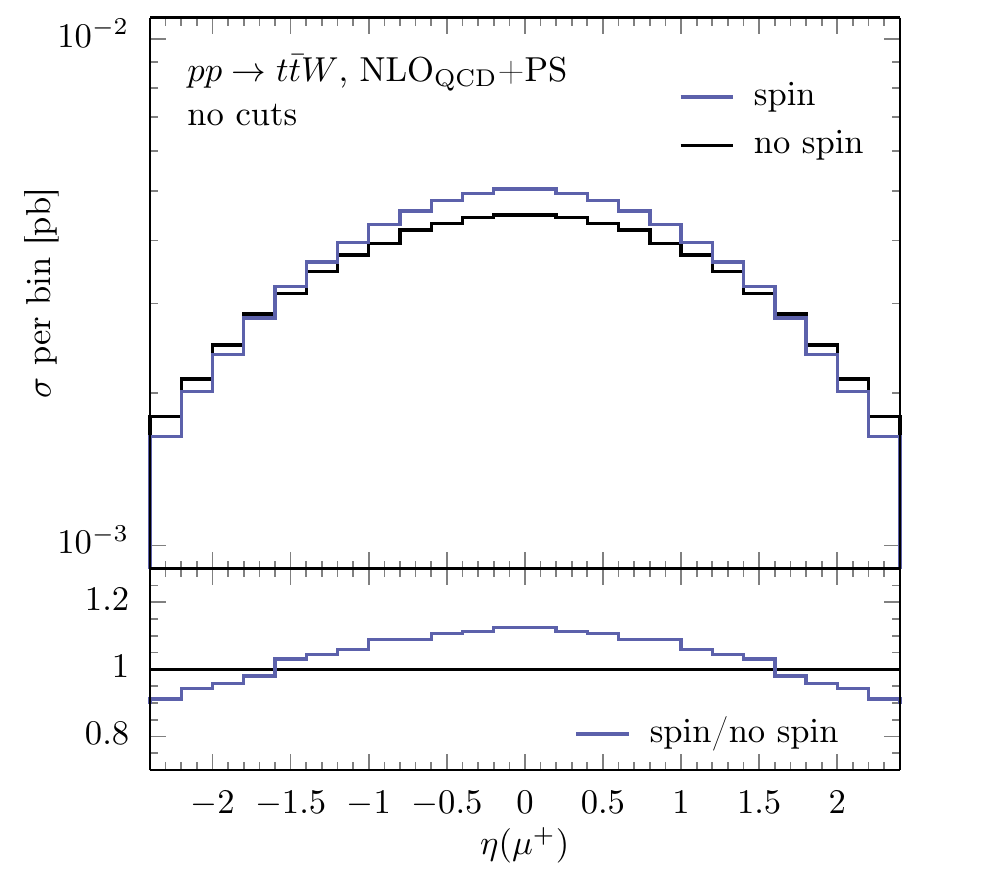}
\includegraphics[width=0.425\textwidth]{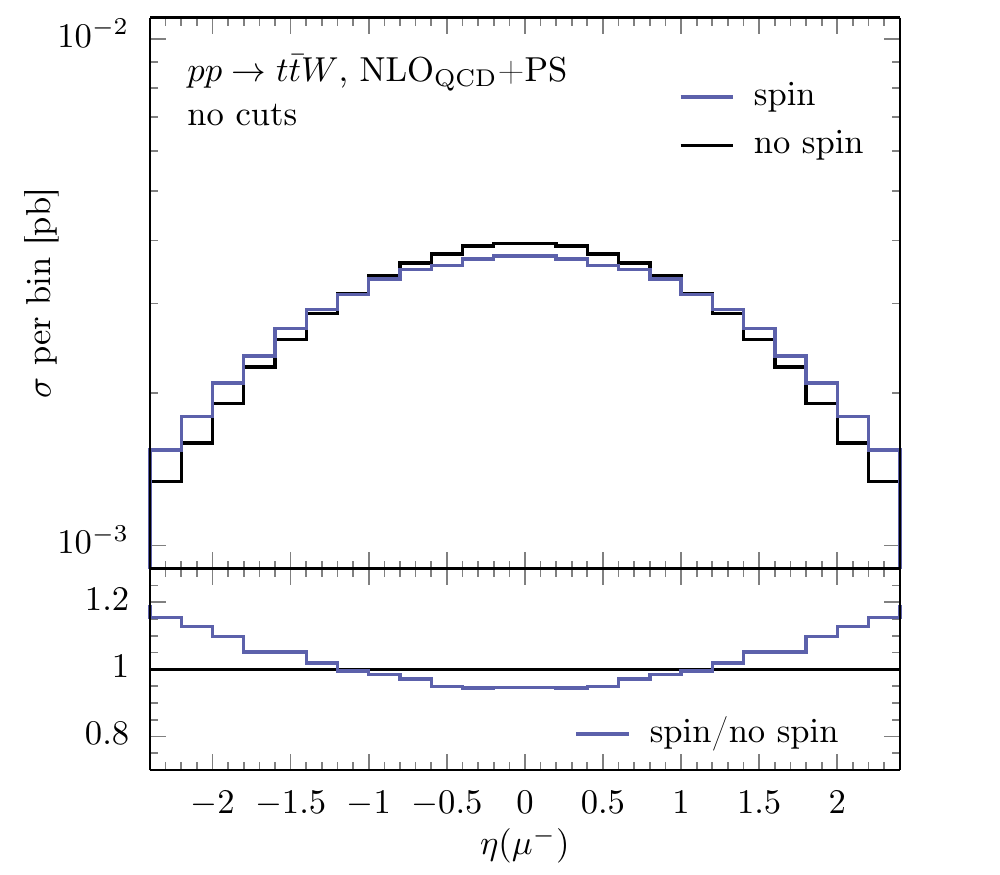}
\includegraphics[width=0.425\textwidth]{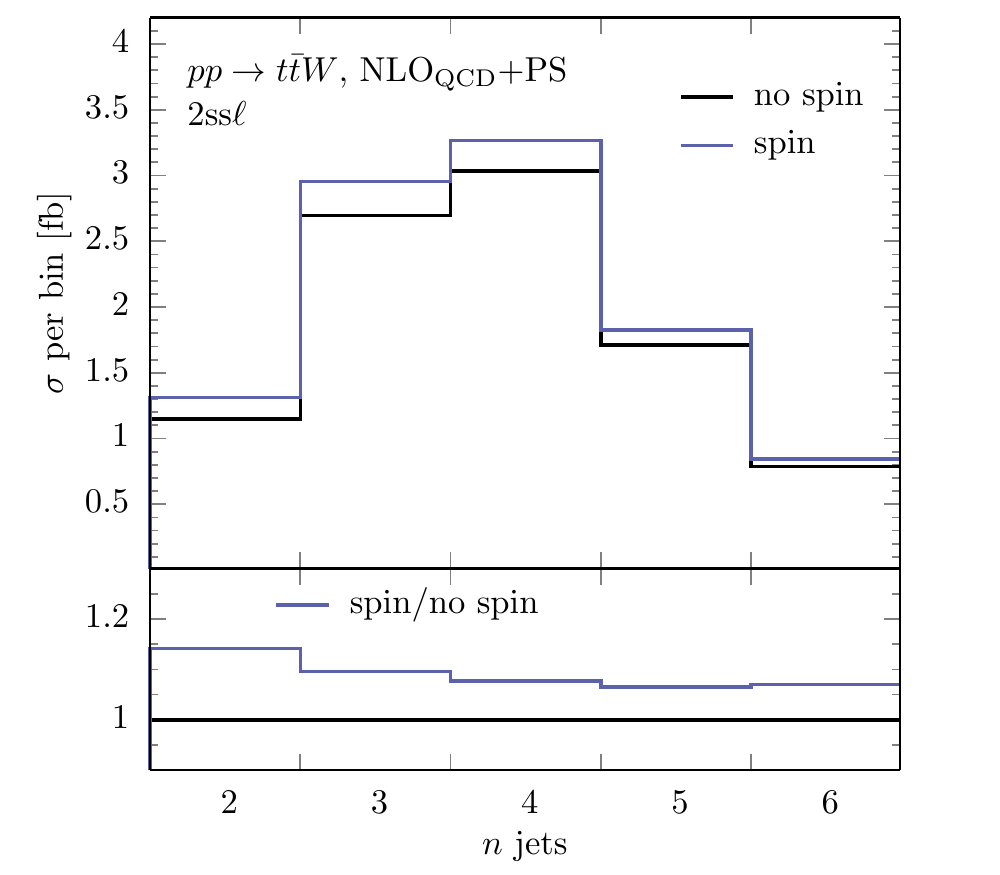}
\includegraphics[width=0.425\textwidth]{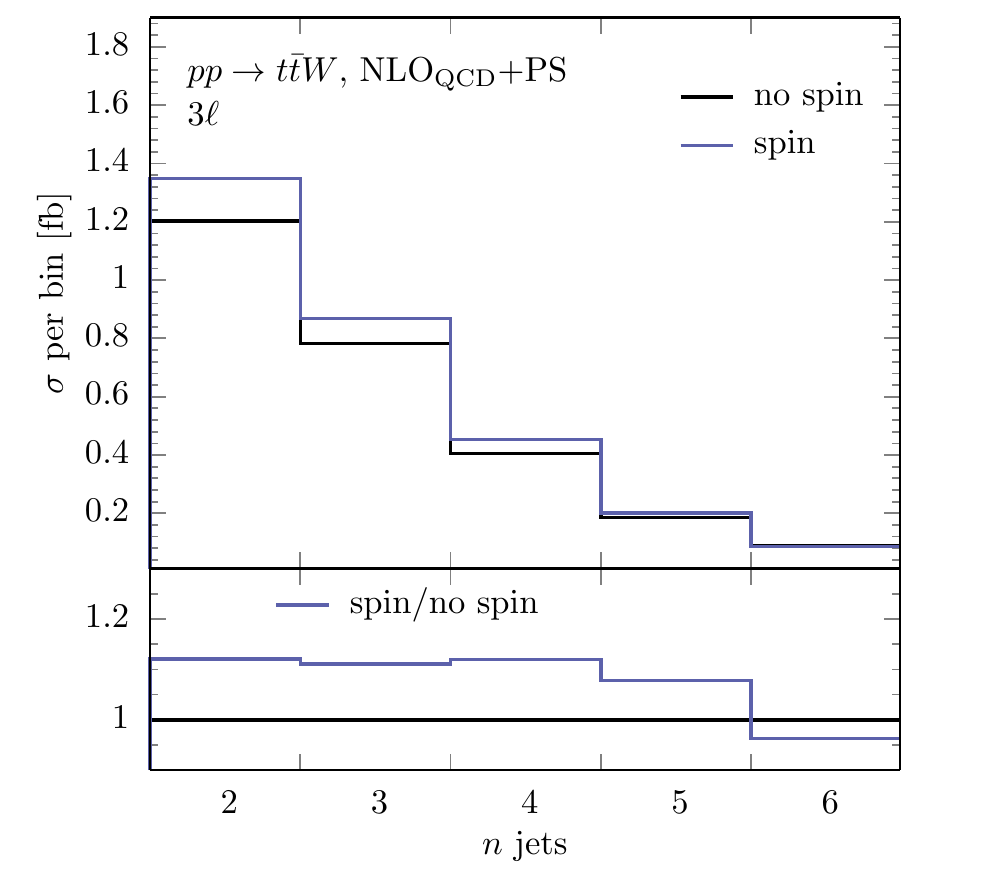}
\caption{Spin correlation effects on lepton pseudo-rapidities and jet multiplicities. The muons in the upper plots are the leading in $p_T$ regardless their origin.}
\label{fig:spin_jets}
\end{figure}

The consequences of these asymmetries on the $t\bar{t}W$ background in
the $t\bar{t}H$ multi-lepton signatures are significant even at the
cross section level. Since the largest of these effects are the spin
correlations, we now compare the effect of these on the inclusive
level as well as on the 2ss$\ell$ and 3$\ell$ signatures.  As a first
step, at the inclusive level (no cuts), we show in the two upper plots
of fig.~\ref{fig:spin_jets} that these effects are not altered by
adopting the realistic setup described in section \ref{sec:calc_par},
i.e.~including QED-shower, hadronisation, and underlying event.
\begin{table}[t!]
\renewcommand{\arraystretch}{1.5}
\begin{center}
\resizebox{\textwidth}{!}{
\begin{tabular}{c @{\qquad} c @{\qquad} c c c c c c c}
\toprule
 Jet multiplicity: & inclusive & 0 & 1 & 2 & 3 & 4 & 5 & 6 \\
\midrule
no cuts & 1.977(2) & 2.88(4) & 2.43(1) & 2.218(7) & 2.087(4) & 2.003(3) & 1.956(3) & 1.916(3) \\
no cuts-no spin & 1.977(1) & 2.90(4) & 2.45(1) & 2.205(7) & 2.087(5) & 2.003(4) & 1.956(3) & 1.920(3) \\
2ss$\ell$ & 1.99(2) & - & - & 2.30(3) & 2.02(2) & 1.96(2) & 1.94(3) & 1.84(4) \\
2ss$\ell$-no spin & 1.84(1) &  &  & 1.90(3) & 1.84(2) & 1.84(2) & 1.84(3) & 1.72(4) \\
3$\ell$ & 1.88(2) & - & - & 1.89(3) & 1.92(4) & 1.81(5) & 1.83(8) & 1.8(1) \\
3$\ell$-no spin & 1.84(2) &  &  & 1.81(3) & 1.82(4) & 1.86(5) & 1.90(8) & 1.9(1) \\
\bottomrule
\end{tabular}
}
\end{center}
\caption{Charge ratio ${\sigma_{t \bar t W^+}}/{\sigma_{t \bar t
      W^-}}$ in different signatures. The scale uncertainties can be
  taken to be correlated and are therefore of the order of the
  statistical error (in parantheses) and are not shown.}
\label{table:ratio} 
\end{table}
\noindent
As one can see from figs.~\ref{fig:asymm} and \ref{fig:assoc} and the
upper two plots of fig.~\ref{fig:spin_jets}, the spin correlations
allow more $\ell^+$'s than $\ell^-$'s within the selection criteria
(eq.~\ref{eq:sel}). This, in combination with the fact that there are
more $\ell^+$'s produced than $\ell^-$'s due to the $t\bar t W^+$
cross section being larger than the $t \bar t W^-$ one by a factor of
$\sim$2, increases the fiducial cross section of $t \bar t W$
production in both the 2ss$\ell$ and 3$\ell$ signatures. This can also
be seen in the two lower plots of fig.~\ref{fig:spin_jets}, where we
show the jet multiplicities with and without the spin-correlation
effects for the 2ss$\ell$ and 3$\ell$ signatures in the left and right
plots, respectively. The increase in the cross section due to the spin
correlations between the top and the anti-top is about 10\% and
slightly larger for the lower-multiplicity bins as compared to the
higher ones.

Alternatively, the effects of the spin correlations can be presented
in the value of the charge ratio ${\sigma_{t \bar t W^+}}/{\sigma_{t
    \bar t W^-}}$ for the various signatures. This is shown in
tab.~\ref{table:ratio}. In this table we show the ratio for the total
cross section and bin by bin for the jet multiplicity in both
signatures. As a reference and in accordance with
fig.~\ref{fig:spin_jets}, we also show the same ratio before any
selections or cuts (no cuts) as well as before the spin-correlation
effects (no spin) are taken into account.  As expected, the inclusive
result (before any selections or cuts) is not affected by the spin
correlations. Without including the latter, in both signatures the
charge ratio decreases from 1.977 to 1.84. This is mostly due to the
decrease of the $\eta(e^+)/\eta(e^-)$ ratio in the central
pseudo-rapidity region due to the PDF effect, as shown in
fig.~\ref{fig:assoc}.  By including the spin-correlation effects the
charge ratio in the 2ss$\ell$ signature increases, and accidentally
agrees within the uncertainties with the inclusive result.  In the
$3\ell$ signature the ratio also increases, but less. This is due to
the strong preference of the 2ss$\ell$ signature to the positively
charged lepton pair as shown in the pseudo-rapidity distributions of
fig.~\ref{fig:asymm} and which we will now elaborate on in more
detail.
 
In the 2ss$\ell$ signature it is more often that the anti-top (as
compared to the top) decays hadronically within this signal
region. This is because $\sigma(t \bar t W^+) > \sigma(t \bar t W^-)$
and the (potential) $\ell^+$ from top is more central than the
(potential) $\ell^-$ from the anti-top due to spin correlations. This
results to a larger increase of the charge ratio due to spin
correlations as compared to the $3\ell$ signature, where all three
massive particles need to decay (semi-)leptonically. Hence, for the
$3\ell$ signature also the more-forward $\ell^-$ from the anti-top
needs to be within the selection criteria, resulting in a smaller
increase due to spin correlations than for the 2ss$\ell$
signature. Besides this, also the `PDF' affects the 2ss$\ell$
signature differently from the 3$\ell$ one. In both these signatures
the associated $W$ boson decays leptonically and the `PDF' effect
described in section \ref{sec:Spin} affects them in the same way. This
is not true for the top-quark pair decay products. In the $3\ell$
signature both the top and anti-top quarks decay semi-leptonically,
therefore no extra asymmetry is induced from `PDF' effect. However in
the case of 2ss$\ell$ signature there is an extra asymmetry induced
due to the fact that for the positively charged lepton pair both
leptons will be on average harder and more forward (emerging from $t
\bar t W^+$) as compared to the negatively charged lepton pair
(emerging from $t \bar t W^-$). Even though this effect is opposite to
the one from the top-quark pair spin correlations, the latter is
always dominant, resulting in a larger charge ratio for the 2ss$\ell$
as compared to the 3$\ell$ signatures.

Concerning the jet multiplicitiy, the charge ratio decreases at the
highest jet multiplicities, where the shower effects become
important. We have checked that these results do not change
significantly once we add the EW$_{\rm sub}$ contributions to the NLO
QCD, as we will explain in section \ref{sec:SubEW}. We have also
verified that already without misidentifications or identification
inefficiencies there is a large migration of events from the 3-lepton
decay mode to the 2ss$\ell$ signature (which is included in our
results). However this effect is
expected to be enhanced in the experimental analyses and there will
also be the inverse migration (2-lepton decay mode to $3\ell$
signature). Therefore the results presented in
fig.~\ref{fig:spin_jets} and tab.~\ref{table:ratio} are expected to be
sensitive to these effects and should be reconsidered with full
detector simulation.

\subsection{Subleading EW contributions}
\label{sec:SubEW}

Having understood in section \ref{sec:Spin} the origin of the lepton asymmetries 
and their impact on the different final signatures we proceed to the discussion on
the EW$_{\rm sub}$ contributions, as defined in equation \ref{eq:blobs}.
\begin{figure}[h]
\centering
\begin{tabular}{m{4.5cm} m{4.5cm} m{4.5cm}}
\includegraphics[width=0.235\textwidth]{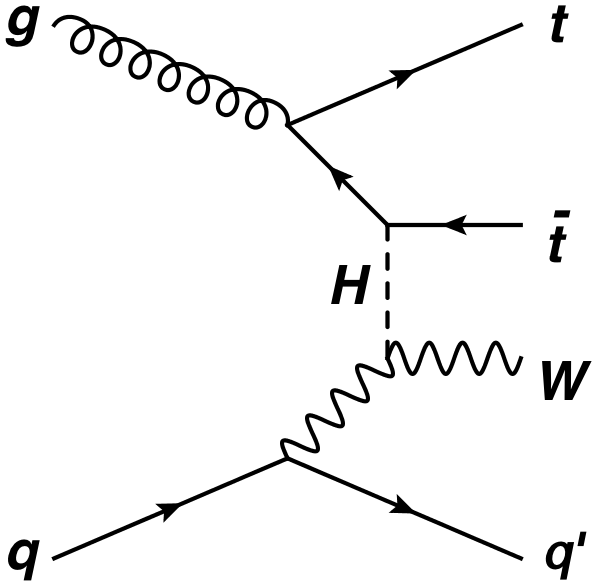} &
\includegraphics[width=0.245\textwidth]{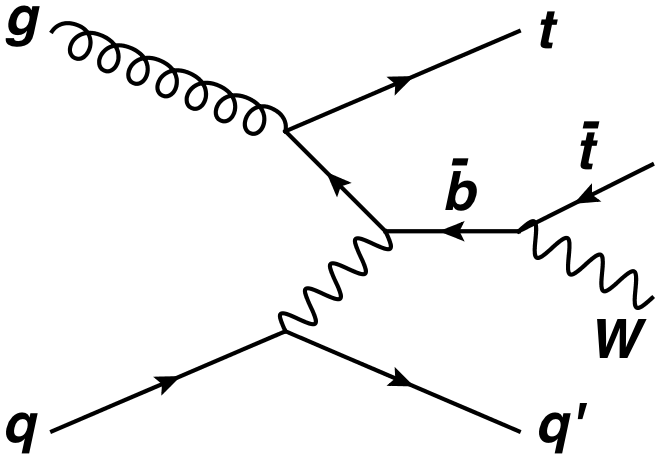} &
\includegraphics[width=0.235\textwidth]{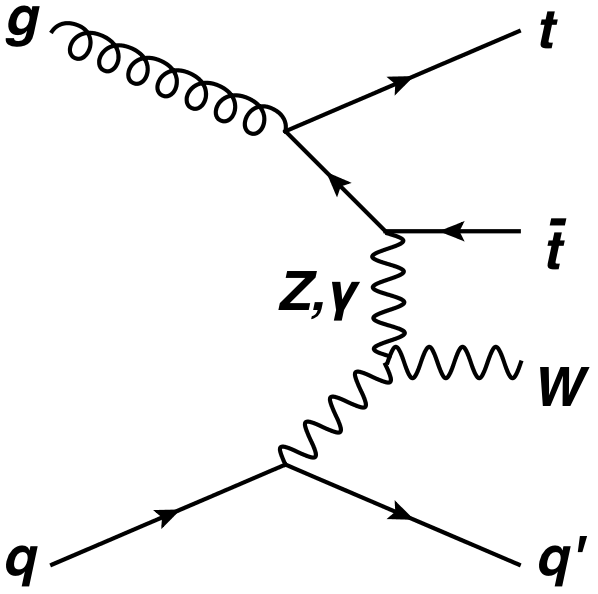}
\end{tabular}
\caption{Dominant representative Feynman diagrams for the EW$_{\rm sub}$ contributions in the $\alpha_s \alpha^3$ perturbative order.}
\label{fig:EW_sub_diagrams}
\end{figure}
\begin{figure}[b]
\centering
\includegraphics[width=0.425\textwidth]{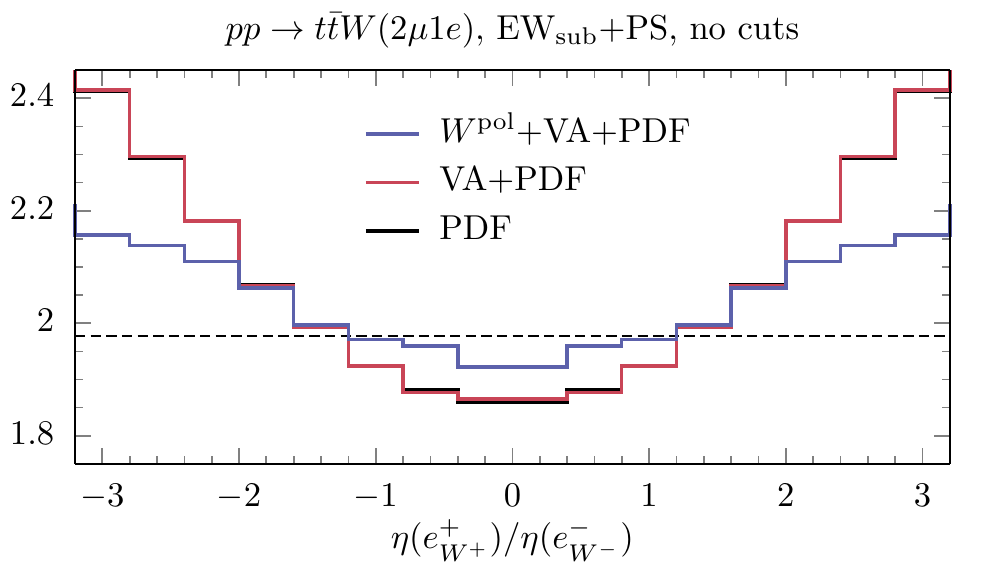}
\includegraphics[width=0.425\textwidth]{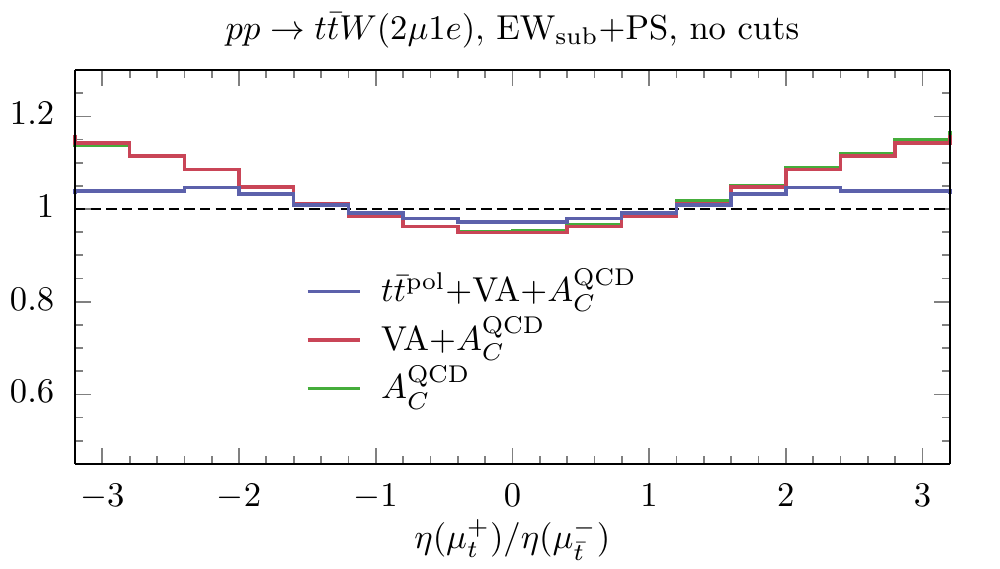}
\caption{Origin of asymmetries in $W$ associated (left) and top-quark pair (right) decay products for the EW$_{\rm sub}$ perturbative orders.}
\label{fig:EWsub_spin}
\end{figure}
\noindent
It is shown in ref.~\cite{Frederix:2018nkq} and discussed in detail in
ref.~\cite{Frederix:2017wme} that the $\sim$10\% the EW$_{\rm sub}$
contributions originate almost exclusively from the $\alpha_s
\alpha^3$ perturbative order (the NLO${}_3$ in the notation of
ref.~\cite{Frederix:2017wme}).  The dominant representative diagrams
of this contribution are shown in
fig.~\ref{fig:EW_sub_diagrams}. These contributions are $qg$ initiated
with different structure w.r.t.~the $q\bar q$ initiated contributions
that cause the large lepton asymmetries already at LO. Therefore the
presence of the $W$ boson does not result to the same spin
correlations for the top pair decay products. This can be seen in the
plots in fig.~\ref{fig:EWsub_spin}, for which we follow the same setup
as for figs.~\ref{fig:asymm} and \ref{fig:assoc}. Regarding the
associated $W$ boson decay products (left plot) the contribution of
the various effects is similar to the corresponding ones at NLO in QCD
(lower left plot in fig.~\ref{fig:assoc}).  However this is not the
case for the top-quark pair decay products (right plot).  Due to the
aforementioned differences of these contributions, the large effect of
the top-quark pair spin correlations (lower right plot in
fig.~\ref{fig:asymm}) is not there.  This last remark shows that the
inclusion of the EW$_{\rm sub}$ contributions slightly flattens the
asymmetries of the top-quark pair leptons. The overall effect is
small, since the total contribution from the EW$_{\rm sub}$ is only
$\sim$10\%. As a result the charge ratio presented in
tab.~\ref{table:ratio} is not altered within the given statistical MC
errors by the inclusion of the EW$_{\rm sub}$ perturbative orders. In
the next paragraph we explore the effect of the EW$_{\rm sub}$
contributions to the total cross section and the jet multiplicities
within the selected signatures.
\begin{figure}[t]
\centering
\includegraphics[width=0.425\textwidth]{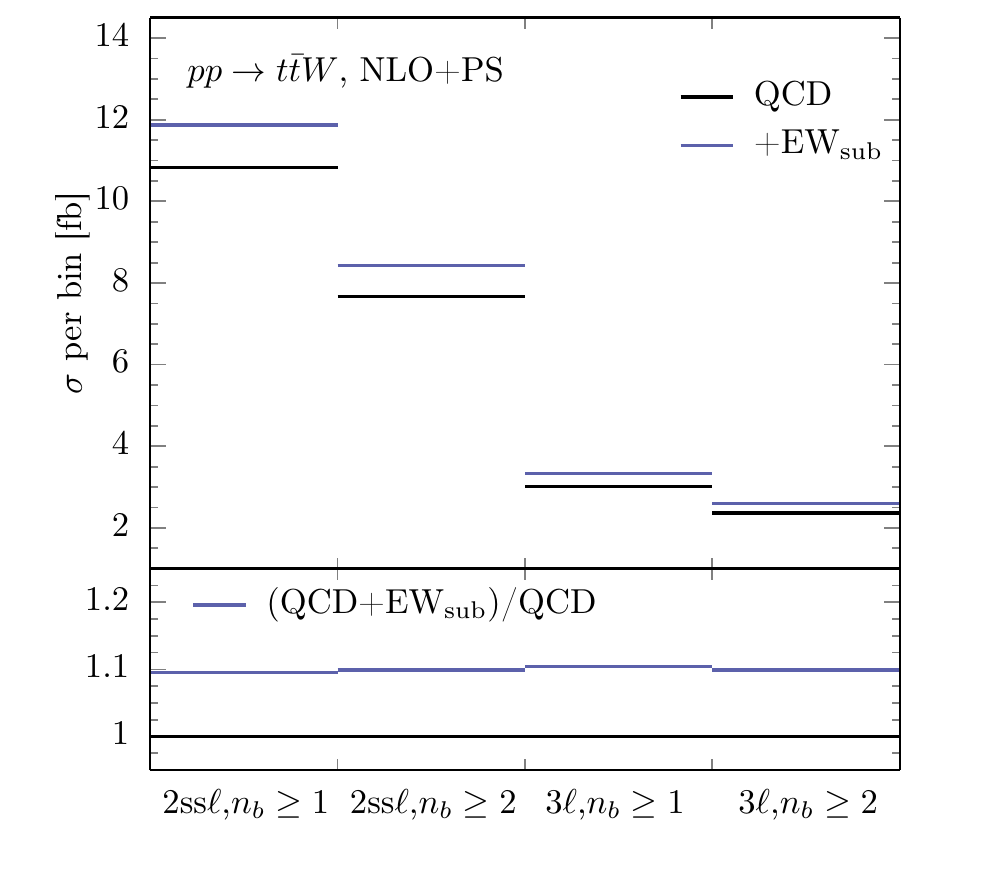} \\
\includegraphics[width=0.425\textwidth]{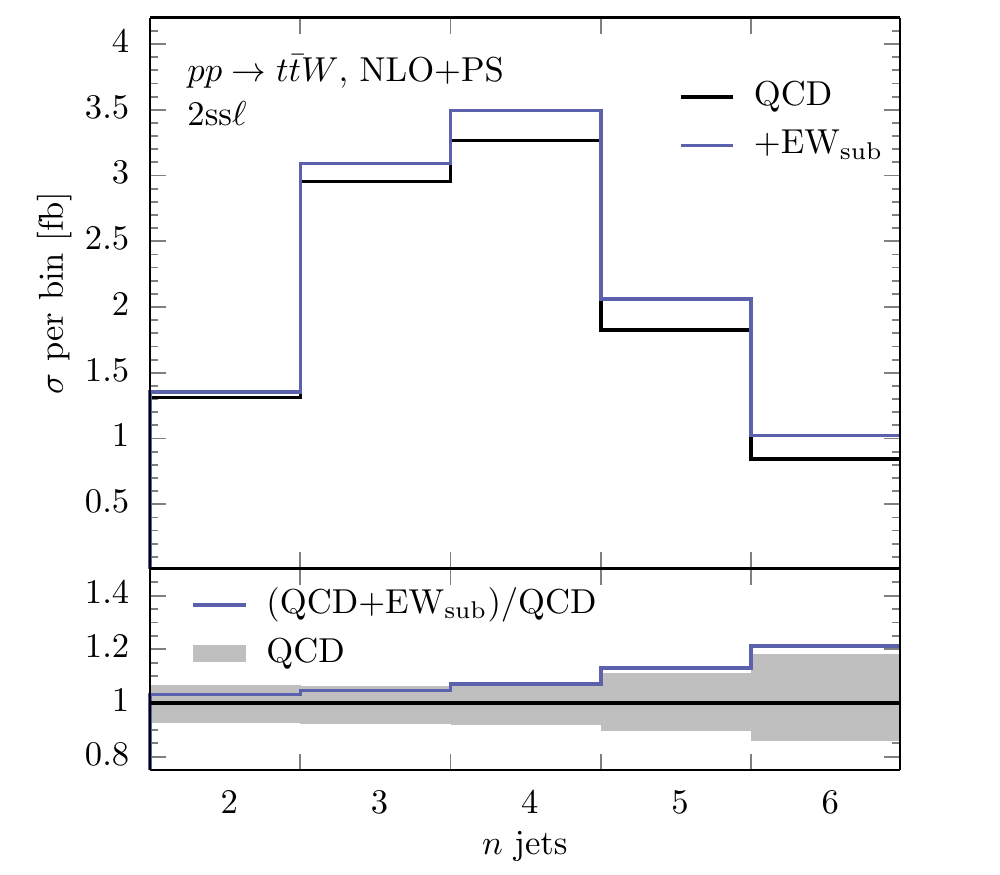}
\includegraphics[width=0.425\textwidth]{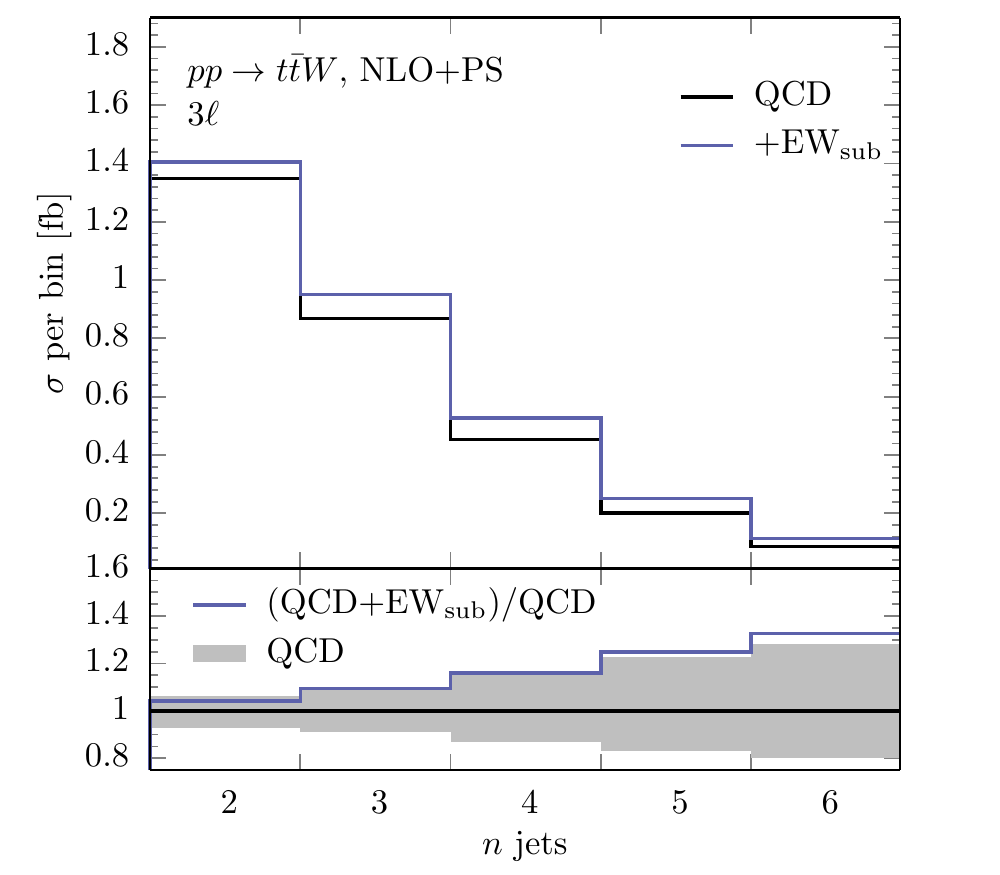}
\caption{Effect of the EW$_{\rm sub}$ contributions on the cross section and the jet multiplicities for the 2ss$\ell$ and $3\ell$ signatures.}
\label{fig:EW_jets}
\end{figure}
\noindent
 
Starting from the NLO QCD production the inclusion of the EW
perturbative orders up to now is done by applying a flat scaling
factor $0.96$ for the $-4\%$ contribution of the $\alpha_s^2 \alpha^2$
perturbative order and a $1.09$ for the contribution of the EW$_{\rm
  sub}$ contributions~\cite{ATLAS:2019nvo}. In the upper plot of
fig.~\ref{fig:EW_jets} we start by showing the effect of the latter on the cross sections for both the 2ss$\ell$ and 3$\ell$ signatures requiring at
least one ($n_b \geq 1$) or two ($n_b \geq 2$) b-jets. As shown in the plot, in
all the selected signatures there is a $10\%$ effect of the EW$_{\rm
  sub}$ contributions in agreement with the aforementioned applied
scaling factor. In the lower two plots of fig.~\ref{fig:EW_jets} the
jet multiplicities in the 2ss$\ell$ and 3$\ell$ are shown in the left
and right plot, respectively. From these plots one can see that the
effects from the EW$_{\rm sub}$ are not flat. In particular, they have
a rather small effect in the low jet-multiplicity bins, but are
significantly larger in the higher jet-multiplicity bins. Hence, they
behave opposite from the spin-correlation effects presented in the lower
plots of fig.~\ref{fig:spin_jets}.  In the lower insets, also the
scale uncertainties (obtained in the usual way by taking the envelope
of the $3 \times 3 = 9$ point variation of the renormalisation and
factorisation scales) in the $t\bar{t}W$ production process are
shown. The EW$_{\rm sub}$ are just at the edge of the scale
uncertainty band\footnote{One should keep in mind that, in particular
  for the larger multiplicities, there is also a significant
  uncertainty coming from the parton shower modeling, which we have
  not included here.}. The main reason for this behaviour at the
differential level is the fact that the dominant topologies of the
EW$_{\rm sub}$ contributions (fig.~\ref{fig:EW_sub_diagrams}) have an
extra parton.  As a result, the peaks of the jet multiplicities (lower
plots of fig.~\ref{fig:spin_jets}) are shifted to the right and
furthermore the extra parton increases the sources of radiation.

\section{Conclusions}
\label{sec:concl}
In this work we discussed two non-negligible effects in $t\bar{t}W$
production: spin correlations in the top-quark pair and the large, formally subleading EW corrections (mainly)
induced by $tW \to tW$ scattering. In the current experimental analyses 
the former is included, whereas the latter is simulated via a flat $K$-factor. 
Concerning the spin correlations we disentangled every source of asymmetry relevant 
to the final signatures. We further studied and presented, for the first time, the 
impact of the EW$_{\rm sub}$ contributions on the selected signatures within a 
realistic analysis setup. We have found that both effects
enhance the $t\bar{t}W$ background in the 2ss$\ell$ and 3$\ell$ signal
regions of $t\bar{t}H$ production by approximately 10\%. However,
their effects are not flat in the phase-space. In particular, we
considered the cross sections binned in jet multiplicity and found
that the spin correlation effects enhance the low jet multiplicities
more than high jet multiplicities, and the EW$_{\rm sub}$ inducing an
opposite effect. However, since the induced differences in shapes are
rather different, also the combined effect of spin correlations and
EW$_{\rm sub}$ contributions is not be flat in phase-space.  Hence,
we can conclude that both effects are important, and that both effects
need to be included in the analysis in a differential manner. For the
latter effect, this work shows, for the first time, that this can
indeed be done: it is possible to use the default MC@NLO matching, as available in Madgraph5\_aMC@NLO, to
include the large EW$_{\rm sub}$ contributions (which include $tW \to
tW$ scattering) within the standard event generation framework.

\section*{Acknowledgments}

This work is done in the context of and supported by the Swedish
Research Council under contract number 2016-05996. IT would like to
thank Rohin Narayan for the detailed discussions on the experimental
analysis.


\end{document}